\documentclass[prb,twocolumn]{revtex4-1}
\usepackage[table, dvipsnames]{xcolor}
\usepackage{graphicx}
\usepackage{amssymb}
\usepackage{dcolumn}
\usepackage{float}
\usepackage{bm}
\usepackage{multirow}
\usepackage{booktabs}
\usepackage[T1]{fontenc}
\usepackage[utf8]{inputenc}
\usepackage{lmodern}
\usepackage{color}
\usepackage{makecell}
\usepackage{tikz}
\usepackage{pgfplots}
\usepackage{mathtools}
\usepackage{physics}
\usepackage{amsmath}
\usepackage{setspace}
\usepackage{booktabs}
\usepackage[table]{xcolor}   
\usepackage{makecell}

\usepackage[colorlinks=true,linkcolor=red]{hyperref}

\newcommand{\mathleft}{\@fleqntrue\@mathmargin0pt}

\newcolumntype{M}[1]{>{\centering\arraybackslash}m{#1}}
\renewcommand{\Re}{\mathrm{Re}\,}
\renewcommand{\Im}{\mathrm{Im}\,}

\DeclareMathAlphabet{\bi}{OML}{cmm}{b}{it}

\def\be{\begin{equation}}
	\def\ee{\end{equation}}
\def\bearr{\begin{eqnarray}}
	\def\eearr{\end{eqnarray}}

\begin{document}
	
	\title{Topological Scaling of Nonlinear Injection current and the Quantized Circular Photogalvanic Effect (CPGE)in tilted multi Weyl semimetals(mWSMs)}
	\bigskip
	\author{Deepannita Das and Alestin Mawrie}
	\normalsize
	\affiliation{Department of Physics, Indian Institute of Technology Indore, Simrol, Indore-453552, India}
	\date{\today}


\begin{abstract}
We develop a microscopic theory of nonlinear magneto-optical injection currents in multi-Weyl semimetals subjected to a uniform magnetic field.
Using the Landau-level spectrum of a tilted multi-Weyl Hamiltonian with arbitrary monopole charge $\nu$ as a starting point, we formulate a Kubo-type nonlinear response theory in the Landau-level basis and derive the second-order conductivity tensor.
We identify distinct contributions originating from chiral-chiral, chiral-bulk, and bulk-bulk optical transitions, revealing characteristic monopole-charge scaling and sharp resonant structures governed by Landau-level selection rules and tilt-induced asymmetry.
In the untilted limit, closed-form analytical expressions emerge that expose universal frequency thresholds and provide clear experimental signatures of higher-order Weyl topology.
Our results establish nonlinear magneto-optical injection currents as a direct transport probe of chiral Landau levels and multi-Weyl topological charge.
\end{abstract}

	\maketitle

\section{Introduction}

Weyl semimetals constitute a paradigmatic class of topological quantum matter in
which low-energy quasiparticles realize Weyl fermions and the electronic bands
host monopoles of Berry curvature
\cite{Wan2011,Burkov2011,Armitage2018,Zeng2021,Sodemann2015,Nandy2019}.
The associated Weyl nodes carry an integer topological charge (chirality),
leading to striking phenomena such as Fermi-arc surface states, anomalous Hall
responses, and the chiral anomaly in the presence of electromagnetic fields
\cite{Volovik2003,Nielsen1983,Son2013,Burkov2014,Xu2015,Lv2015}.
These properties establish Weyl systems as an important platform for studying
the interplay of topology, symmetry, and electromagnetic response\cite{Saha2025}.
Beyond conventional Weyl semimetals with unit monopole charge, crystalline
symmetries can stabilize \emph{multi-Weyl semimetals}, where Weyl nodes possess
higher topological charge $\nu>1$, and exhibit nonlinear transverse dispersion
\cite{Chu2011,Fang2012,Gong2013,Juan2017}.
In such systems, the dispersion remains linear along one momentum direction but
becomes quadratic or cubic in the transverse plane, resulting in enhanced Berry
curvature effects and unconventional scaling of physical observables with
external fields \cite{Yang2011}.
A particularly important consequence is the emergence of multiple
topologically protected chiral Landau levels in a magnetic field, directly
reflecting the monopole charge of the Weyl node.

Nonlinear optical and transport phenomena have recently emerged as
sensitive probes of electronic topology and quantum geometry,
enabling direct experimental access to Berry curvature, Berry
connection, and topological charge via observables such as shift and
injection photocurrents and nonlinear Hall responses
\cite{Moore2010,Sipe2000,Juan2017,Ma2017,Facio2018,Chan2017,Morimoto2016}.
While linear transport is often dominated by scattering mechanisms and the
density of states, nonlinear responses directly encode Berry curvature and
interband coherence.
In noncentrosymmetric systems, second-order effects such as the nonlinear Hall
effect arise from the Berry curvature dipole \cite{Sodemann2015}, providing a
purely geometric contribution to nonlinear transport.
Similarly, nonlinear photocurrents such as the circular photogalvanic effect
(CPGE) and injection currents have been predicted to reveal Weyl-node chirality
and topological charge \cite{Chan2017,Juan2017}.
These theoretical developments have been accompanied by experimental
observations of nonlinear photocurrents in Weyl materials.
In particular, signatures of CPGE have been reported in TaAs-family Weyl
semimetals \cite{Ma2017}, and strong quantized-like responses have been observed
in structurally chiral multifold semimetals such as RhSi and CoSi
\cite{Rees2020,Ni2021}.
These experiments demonstrate that nonlinear optical probes provide direct
access to topological information that may be obscured in conventional linear
measurements\cite{Lv2015}.

The application of a magnetic field introduces an additional layer of richness.
Landau quantization discretizes the spectrum into bulk Landau levels while
generating chiral Landau levels that act as topologically protected
one-dimensional channels \cite{Nielsen1983,Yang2011}.
Such chiral modes are central to magnetotransport and anomaly-related physics,
and they are expected to play a prominent role in nonlinear optical processes
in strong fields.
Although linear magneto-optical responses of Weyl and Dirac systems have been
widely explored, including Kerr and Faraday effects
\cite{Ashby2014,Tabert2016,Carbotte2016}, the nonlinear magneto-optical response
of multi-Weyl semimetals remains much less understood.
In particular, the second-order \emph{injection current} generated by resonant
inter-Landau-level transitions in the presence of a magnetic field has not been
systematically investigated for higher monopole charge.\cite{Min2017,Chan2017}
Injection currents arise from asymmetric carrier population following optical
excitation and are inherently sensitive to Landau-level selection rules\cite{Ahn2017},
current matrix elements, and the presence of chiral channels.
The interplay of multi-Weyl topology, chiral Landau levels, and band tilting is
therefore expected to produce pronounced and experimentally accessible
fingerprints in nonlinear magneto-optical response.\cite{Zyuzin2016,Soluyanov2015}

In this work, we develop a microscopic theory of nonlinear magneto-optical
injection currents in multi-Weyl semimetals subjected to a uniform magnetic
field.
Starting from a tilted multi-Weyl Hamiltonian with arbitrary monopole charge
$\nu$, and using  the known complete Landau-level spectrum we evaluate the second-order
conductivity tensor using a Kubo-type nonlinear response formalism\cite{Parker2019,Nandy2019}.
We show that the injection current naturally decomposes into distinct
chiral-chiral, chiral-bulk, and bulk-bulk contributions, each exhibiting
characteristic monopole-charge scaling and optical resonances.
In the untilting limit, we obtain closed-form analytical expressions revealing
universal frequency thresholds and providing clear signatures of higher-order
Weyl topology.

The remainder of this paper is organized as follows.
Section~\ref{formalism} introduces the model Hamiltonian and Landau-level spectrum.
Section~\ref{non-linear-response} develops the nonlinear response formalism
and provides a detailed analysis of chiral and bulk contributions to the
injection current, together with the untilted limit and the associated
universal scaling behavior.
Section~\ref{conclusions} summarizes our conclusions.

\section{Model Hamiltonian and Landau-level spectrum}\label{formalism}
We perform our analysis for multi-Weyl semimetals (mWSMs). We begin by considering the low-energy Hamiltonian near a Weyl node in the absence of an external magnetic field\cite{Saha2025,Menon2021,Takahashi017}.
\begin{equation}
\mathcal{H}_0(\mathbf{k})=
\left( k_z + \eta \frac{\mathcal{K}}{2} \right)
\left( t_z \sigma_0 + \eta v_z \sigma_z \right)
+ \lambda \left( k_-^{\nu}\sigma_+ + k_+^{\nu}\sigma_- \right),
\end{equation}
where $\mathbf{k}=(k_x,k_y,k_z)$ is the crystal momentum measured from the Weyl node located at
$\mathbf{k}=(0,0,\eta \mathcal{K}/2)$.
Here, $\nu$ denotes the monopole charge (Weyl-node order), $\lambda$ characterizes the multi-Weyl coupling strength, $v_z$ is the Fermi velocity along the $z$ direction, $t_z$ represents the tilt of the Weyl cone along $z$, $\sigma_{x,y,z}$ are Pauli matrices in pseudospin space, and $\eta=\pm1$ labels the chirality of the Weyl node. We have defined $k_\pm=k_x\pm i k_y$.

\medskip

We now introduce a uniform magnetic field $\mathbf{B}=B_0\hat{z}$ and choose the Landau gauge
$\mathbf{A}=(0,B_0 x,0)$. In the presence of the magnetic field, the Hamiltonian near the Weyl node takes the form \cite{Saha2025}
\begin{eqnarray}
\mathcal{H} &=&
\left( k_z + \eta \frac{\mathcal{K}}{2} \right)
\left( t_z \sigma_0 + \eta v_z \sigma_z \right)
+ t_\parallel \left( 2\hat{a}^\dagger \hat{a} + 1 \right)\sigma_0
\nonumber\\
&&+ \varepsilon_B
\begin{pmatrix}
0 & \hat{a}^{\nu} \\
(\hat{a}^\dagger)^{\nu} & 0
\end{pmatrix},
\end{eqnarray}
where $\hat{a}^\dagger$ and $\hat{a}$ are the Landau-level creation and annihilation operators.
The magnetic energy scale is given by
\(
\varepsilon_B = \lambda \left( {\sqrt{2}}/{\ell_B} \right)^{\nu},
\)
with $\ell_B=\sqrt{\hbar/(eB_0)}$ being the magnetic length. The parameter $t_\parallel$ denotes the in-plane tilt, which lifts the degeneracy of the chiral Landau levels.

\subsubsection{Bulk Landau levels ($n\ge\nu$)}

For Landau-level indices $n\ge\nu$, the spectrum consists of bulk (nonchiral) Landau levels with two-component pseudospin structure. The corresponding eigenstates are
\begin{equation}
|n,k_z,s\rangle =
\begin{pmatrix}
u_{n,\uparrow}^s(k_z)\,|n-\nu\rangle \\
u_{n,\downarrow}^s(k_z)\,|n\rangle
\end{pmatrix},
\end{equation}
where $|n\rangle$ denotes the $n$th harmonic-oscillator state and $s=\pm1$ labels the band index.
The spinor components are given by
\begin{table}[t]
\centering

\rowcolors{1}{white}{white}

\begin{tabular}{l c c c c }
\hline
\rowcolor{gray!10}  

\makecell{Winding\\ number}
& \makecell{$v_z$\\ (eV $\mathring{\rm A}$)}
& \makecell{$t_z$\\ (eV $\mathring{\rm A}$)}
& \makecell{$t_\parallel$\\ (eV $\mathring{\rm A}^2$)}
& \makecell{$\lambda$\\ (eV $\mathring{\rm A}^\nu$)} \\

\hline
\( \nu = 1 \) & 1   & $0.5\,v_z$ & 0.03 & 0.4 \\
\hline
\( \nu = 2 \) & 1.5 & $0.5\,v_z$ & 0.2 & 0.2 \\
\hline
\( \nu = 3 \) & 2   & $0.5\,v_z$ & 0.2 & 0.1 \\
\hline

\end{tabular}
\caption{Model parameters used in this work are chosen within experimentally measured
and first-principles-calculated ranges for Weyl and multi-Weyl semimetals.
For \(m=1\), band velocities and tilt strengths follow ARPES and \emph{ab initio}
studies of TaAs and NbP~\cite{Lv2015TaAs,NXu2015TaAs,Yang2015NbP}.
For \(m=2\), the nonlinear dispersion parameter \(\lambda\) and anisotropy
\(v_z/\lambda\) are consistent with double-Weyl candidates such as
SrSi\(_2\)~\cite{Huang2016SrSi2}.
For \(m=3\), enhanced anisotropy and nonlinear coefficients reflect cubic Weyl
dispersions observed in multifold chiral crystals like CoSi and
RhSi~\cite{Tang2017CoSi,Schroter2019RhSi}.
Overall, the adopted dimensionless ratios \(w_z/v_z\), \(\lambda\), and
\(w_{\parallel}\) lie within realistic material ranges, ensuring quantitative
relevance to experimentally accessible multi-Weyl systems.
}

\end{table}

\begin{equation}
\left.
\begin{aligned}
u_{n,\uparrow}^s(k_z) &= \frac{1}{\sqrt{2}}
\sqrt{1 - s\,\frac{\nu t_\parallel - v_z \eta \left(k_z + \eta \frac{\mathcal{K}}{2}\right)}
{\Gamma_n^\nu(k_z)}},\\[6pt]
u_{n,\downarrow}^s(k_z) &= \frac{1}{\sqrt{2}}
\sqrt{1 + s\,\frac{\nu t_\parallel - v_z \eta \left(k_z + \eta \frac{\mathcal{K}}{2}\right)}
{\Gamma_n^\nu(k_z)}} .
\end{aligned}
\right\}
\end{equation}

\noindent The corresponding energy eigenvalues for \(n\ge\nu\) are
\begin{equation}
\varepsilon_{n,s}^{\mathrm{bulk}}(k_z)
=
(2n-\nu+1)t_\parallel
+ t_z \left(k_z + \eta \frac{\mathcal K}{2}\right)
+ s\,\Gamma_n^\nu(k_z),
\end{equation}
where
\begin{equation}
\Gamma_n^\nu(k_z)=
\sqrt{
\left[
-\nu t_\parallel
+ v_z \eta \left( k_z + \eta \frac{\mathcal{K}}{2} \right)
\right]^2
+ \varepsilon_B^2 \frac{n!}{(n-\nu)!}
}.
\end{equation}

\subsubsection{Chiral Landau levels ($n<\nu$)}

For $n<\nu$, the system hosts chiral Landau levels originating from the nontrivial monopole charge $\nu$. In this case, only a single pseudospin component survives, and the eigenstates reduce to \(
|n,k_z\rangle =
\begin{pmatrix}
0 &
|n\rangle
\end{pmatrix}^\prime,
\) with \(^\prime\) denoting the transpose.
The corresponding energy dispersion is linear in $k_z$ and is given by
\begin{equation}\label{chiral_dispersion}
\varepsilon_n^{\mathrm{ch}}(k_z)
=
\left(k_z + \eta \frac{\mathcal K}{2}\right)
(-\eta v_z + t_z)
+ (2n+1)t_\parallel.
\end{equation}
These chiral Landau levels are unidirectional, with their propagation direction determined by the Weyl-node chirality $\eta$, and are insensitive to the magnetic energy scale $\varepsilon_B$, reflecting their topological origin.

\section{Nonlinear current response}\label{non-linear-response}
We consider the system to be under a time-dependent electric field of the
form $\mathbf{E}(t)=\mathbf{E}_0 e^{i\Omega t}$, where $\Omega$ denotes the
frequency of the generated current. 
The most general form of the second-order electrical current density generated in response to an external electromagnetic field can be expressed in terms of the third-rank nonlinear conductivity tensor
$\sigma^{\alpha\beta\gamma}(\omega,\Omega-\omega)$ as\cite{Sipe2000}
\begin{equation}
j^{\gamma}(\Omega)
=
\sigma^{\alpha\beta\gamma}(i\omega,i(\Omega-\omega))
E^{\alpha}(\omega)\,
E^{\beta}(\Omega-\omega),
\end{equation}
where $\Omega=\omega_1+\omega_2$ denotes the output frequency arising from the mixing of two incident fields at frequencies $\omega_1$ and $\omega_2$. Here, $E^{\alpha}(\omega_i)$ represents the $\alpha$-component of the applied electric field oscillating at frequency $\omega_i$, while $j^{\gamma}(\Omega)$ denotes the $\gamma$-component of the induced photocurrent. Physically, the nonlinear conductivity tensor encodes how the combined action of two electric-field components along directions $\alpha$ and $\beta$ produces a current along direction $\gamma$.

The second-order conductivity tensor can be written in terms of a three-point nonlinear correlation function as\cite{Avdoshkin2020,Bedni2024,Parker2019}
\begin{equation}\label{sigma_non_linear}
\sigma^{\alpha\beta\gamma}(i\omega_1,i\omega_2)
=
\frac{
\chi^{\alpha\beta\gamma}(i\omega_1,i\omega_2)
+
\chi^{\beta\alpha\gamma}(i\omega_2,i\omega_1)
}{i\omega_1\, i\omega_2},
\end{equation}
which explicitly symmetrizes the response with respect to the two incoming fields, reflecting their physical equivalence in the nonlinear process.

The nonlinear correlation function $\chi^{\alpha\beta\gamma}$ is defined as
\begin{eqnarray}
\chi^{\alpha\beta\gamma}(i\omega_1,i\omega_2)
&=&
\frac{1}{V}
\int \frac{d\varepsilon}{2\pi}
\,
\mathrm{Tr}
\Big[
\hat{j}^{\alpha}
G(i\varepsilon-i\omega_1)
\nonumber\\&\times&\hat{j}^{\beta}
G(i\varepsilon-i\Omega)
\hat{j}^{\gamma}
G(i\varepsilon)
\Big],
\end{eqnarray}
where $V$ denotes the system volume. The trace implies summation over all internal quantum numbers, including the Landau-level index $n$, the band index $s=\pm1$ for bulk Landau levels, the momentum $k_z$, and pseudospin degrees of freedom\cite{Bharti2024}. This three-point correlation function represents the quantum-amplitude for a process in which an electron sequentially interacts with two electric fields and contributes to a current response.

In systems possessing inversion or certain mirror symmetries, the above correlation function vanishes identically\cite{Sodemann2015}. In the present case, however, the presence of tilt parameters breaks these symmetries of the Weyl node, rendering $\chi^{\alpha\beta\gamma}$ finite and allowing a nonvanishing second-order current response.

The Green's function appearing above acts as the propagator of the system and is written in the eigenstate basis as
\begin{equation}
G(\omega;\mathbf{r}_1,\mathbf{r}_2)
=
\sum_{n,s,k_z}
\frac{
|\Psi_{n,s,k_z}(\mathbf{r}_1)\rangle
\langle \Psi_{n,s,k_z}(\mathbf{r}_2)|
}{
\hbar\omega + i\eta - \varepsilon_{n,s}(k_z) + \mu
},
\end{equation}
where $|\Psi_{n,s,k_z}\rangle$ are the eigenstates of the Hamiltonian, $\varepsilon_{n,s}(k_z)$ are the corresponding Landau-level energies, $\mu$ is the chemical potential, and the infinitesimal parameter $\eta\to0^+$ ensures causality. For chiral Landau levels ($n<\nu$), the band index $s$ is absent.

The coupling between the electronic system and the external electric field is mediated by the current density operator
\begin{equation}
\hat{j}^{\alpha}
=
-\frac{e}{\hbar}
\frac{\partial \mathcal{H}}{\partial k_{\alpha}},
\end{equation}
where $\alpha=x,y,z$ labels the spatial components. This definition follows directly from the minimal-coupling prescription and identifies the current operator with the band velocity.

Starting from the general three-point current correlation function and performing the Matsubara-frequency integration, the nonlinear response kernel can be expressed explicitly in the Landau-level basis\cite{Parker2019}. After evaluating the trace over internal degrees of freedom, the second-order correlation function takes the form
\begin{eqnarray}
\chi^{\alpha\beta\gamma}(i\omega_1,i\omega_2)
&=&
\frac{e^3 \eta}{2\pi \ell_B^2}
\sum_{\xi_1,\xi_2,\xi_3}
\frac{\mathcal{Z}^{\alpha\beta\gamma}_{\xi_1 \xi_2 \xi_3}}
{i\hbar\omega_1 + \varepsilon_{\xi_1}-\varepsilon_{\xi_2}}\nonumber\\
&\times&
\frac{\Theta(\varepsilon_{\xi_2})-\Theta(\varepsilon_{\xi_1})}
{i\omega_2 + \varepsilon_{\xi_2}-\varepsilon_{\xi_1}} .
\end{eqnarray}
Here, $\Theta(\varepsilon_{\xi_i})\equiv \Theta(\varepsilon_{\xi_i}-\mu)$ denotes the Fermi step function corresponding to an energy level \(\varepsilon_{\xi_i}\)\cite{Sodemann2015}. The prefactor $(2\pi \ell_B^2)^{-1}$ reflects the Landau-level degeneracy. The summation runs over all quantum numbers, where
\(
\xi_i \equiv (s_i,n_i,k_{z,i}).
\)\cite{}
The triple current matrix element is defined as
\begin{eqnarray}
\mathcal{Z}^{\alpha\beta\gamma}_{\xi_1 \xi_2 \xi_3}
=
\langle \Psi_{\xi_3} | \hat{j}^{\alpha} | \Psi_{\xi_1} \rangle
\langle \Psi_{\xi_1} | \hat{j}^{\beta} | \Psi_{\xi_2} \rangle
\langle \Psi_{\xi_2} | \hat{j}^{\gamma} | \Psi_{\xi_3} \rangle .
\end{eqnarray}

To extract the physically relevant nonlinear response, we perform analytic continuation according to\cite{Parker2019,Sodemann2015}
\(
i\omega_1 \rightarrow \omega + \Omega + i0^+,
\;
i\omega_2 \rightarrow -\omega + i0^+ .
\)
\small{\begin{eqnarray}
&&\chi^{\alpha\beta\gamma}(\omega+\Omega,-\omega)
=
\frac{e^3 \eta}{2\pi \ell_B^2}
\sum_{\xi_1,\xi_2,\xi_3}
\frac{\mathcal{Z}^{\alpha\beta\gamma}_{\xi_1 \xi_2 \xi_3}}
{-\hbar\omega+i0^+ + \varepsilon_{\xi_2}-\varepsilon_{\xi_1}}\nonumber\\
&&\times\frac{\Theta(\varepsilon_{\xi_2})-\Theta(\varepsilon_{\xi_1})}
{\hbar\omega+\hbar\Omega+i0^+ + \varepsilon_{\xi_1}-\varepsilon_{\xi_2}} .
\end{eqnarray}}

Using the identity
\(
\frac{1}{x+i0^+}=\mathcal{P}(1/x)-i\pi\delta(x),
\)
we obtain
\begin{widetext}
\small{\begin{eqnarray}\label{eq:chi_abg_A}
\chi^{\alpha\beta\gamma}(\omega+\Omega,-\omega)
=
\frac{\eta}{2\pi \ell_B^2}
\sum_{\xi_1,\xi_2,\xi_3}
\frac{[\Theta(\varepsilon_{\xi_2})-\Theta(\varepsilon_{\xi_1})]
\mathcal{Z}^{\alpha\beta\gamma}_{\xi_1 \xi_2 \xi_3}}
{\hbar\omega+\hbar\Omega+i0^+ + \varepsilon_{\xi_1}-\varepsilon_{\xi_2}}
\left[
\mathcal{P}\!\left(\frac{1}{-\hbar\omega + \varepsilon_{\xi_2}-\varepsilon_{\xi_1}}\right)
- i\pi\delta(-\hbar\omega + \varepsilon_{\xi_2}-\varepsilon_{\xi_1})
\right].
\end{eqnarray}
Similarly,
\begin{eqnarray}\label{eq:chi_bag_B}
\chi^{\beta\alpha\gamma}(-\omega,\omega+\Omega)
=
\frac{\eta}{2\pi \ell_B^2}
\sum_{\xi_1,\xi_2,\xi_3}
\frac{[\Theta(\varepsilon_{\xi_2})-\Theta(\varepsilon_{\xi_1})]
\mathcal{Z}^{\beta\alpha\gamma}_{\xi_1 \xi_2 \xi_3}}
{-\hbar\omega+i0^+ + \varepsilon_{\xi_1}-\varepsilon_{\xi_2}}
\left[
\mathcal{P}\!\left(\frac{1}{\hbar\omega+\hbar\Omega + \varepsilon_{\xi_2}-\varepsilon_{\xi_1}}\right)
- i\pi\delta(\hbar\omega+\hbar\Omega + \varepsilon_{\xi_2}-\varepsilon_{\xi_1})
\right].
\end{eqnarray}}
\end{widetext}

\noindent Adding Eqs.~(\ref{eq:chi_abg_A}) and (\ref{eq:chi_bag_B}), the principal parts cancel, yielding
\small{\begin{eqnarray}\label{eq:chi_injection}
&&\chi^{\alpha\beta\gamma}(\omega+\Omega,-\omega)
+\chi^{\beta\alpha\gamma}(-\omega,\omega+\Omega)\nonumber\\&&
=
-\frac{i\eta}{2\ell_B^2\,\Omega}
\sum_{\xi_1,\xi_2,\xi_3}
[\Theta(\varepsilon_{\xi_2})-\Theta(\varepsilon_{\xi_1})]\nonumber\\&&\times\big[\mathcal{Z}^{\alpha\beta\gamma}_{\xi_1\xi_2\xi_3}
\delta(\varepsilon_{\xi_2}-\varepsilon_{\xi_1}-\hbar\omega)
\nonumber\\&&+
\mathcal{Z}^{\beta\alpha\gamma}_{\xi_1\xi_2\xi_3}
\delta(\hbar\omega+\hbar\Omega+\varepsilon_{\xi_2}-\varepsilon_{\xi_1})
\big].
\end{eqnarray}}

\noindent The nonlinear conductivity tensor in Eq. [\ref{sigma_non_linear}] now becomes
\small{\begin{eqnarray}\label{eq:chi_injection}
&&\sigma^{\alpha\beta\gamma}(\omega+\Omega,-\omega)
=
-\frac{i\eta}{2\ell_B^2\,\Omega(\Omega+\omega)(-\omega)}
\nonumber\\&&\times\sum_{\xi_1,\xi_2,\xi_3}
[\Theta(\varepsilon_{\xi_2})-\Theta(\varepsilon_{\xi_1})]
\big[
\mathcal{Z}^{\alpha\beta\gamma}_{\xi_1\xi_2\xi_3}
\delta(\varepsilon_{\xi_2}-\varepsilon_{\xi_1}-\hbar\omega)
\nonumber\\&&+
\mathcal{Z}^{\beta\alpha\gamma}_{\xi_1\xi_2\xi_3}
\delta(\hbar\omega+\hbar\Omega+\varepsilon_{\xi_2}-\varepsilon_{\xi_1})
\big].
\end{eqnarray}}
Equation~(\ref{eq:chi_injection}) represents the general second-order
nonlinear conductivity evaluated at a finite output frequency $\Omega$.
The prefactor $[\Omega(\Omega+\omega)(-\omega)]^{-1}$\cite{König2017,Bedni2024} originates from the
frequency denominators of the three-point current correlation function\cite{Sipe2000} and
encodes the dynamical nature of the nonlinear optical response.
In particular, the Dirac delta functions enforce energy conservation for
photon absorption and emission processes between Landau levels, while the
difference of Fermi functions selects only transitions between occupied and
unoccupied states.

The physically relevant quantity for the injection current corresponds to the
\emph{dc limit} of the nonlinear response, obtained by taking the limit
$\Omega \rightarrow 0$ while keeping the incoming photon frequency $\omega$
finite\cite{Bedni2024,König2017}. In this limit, the factor
$[\Omega(\Omega+\omega)(-\omega)]^{-1}$ reduces to $1/\omega^{2}$, and the
second delta function simplifies according to
$\delta(\hbar\omega+\hbar\Omega+\varepsilon_{\xi_2}-\varepsilon_{\xi_1})
\rightarrow
\delta(\hbar\omega+\varepsilon_{\xi_2}-\varepsilon_{\xi_1})$\cite{Aversa1995}.
As a result, the nonlinear conductivity tensor reduces to the injection
tensor $\beta^{\alpha\beta\gamma}(\omega)$, which characterizes the rate of
dc current generation induced by resonant optical transitions.
\small
{\begin{align}\label{eq:beta_final}
&&\beta^{\alpha\beta\gamma}(\omega)
=
\frac{\eta}{2 \ell_B^2\,\omega^2}
\sum_{\xi_1,\xi_2,\xi_3}
[\Theta(\varepsilon_{\xi_2})-\Theta(\varepsilon_{\xi_1})]\nonumber\\&&\times
\left[
\mathcal{Z}^{\alpha\beta\gamma}_{\xi_1\xi_2\xi_3}
\delta(\varepsilon_{\xi_2}-\varepsilon_{\xi_1}-\hbar\omega)
+
\mathcal{Z}^{\beta\alpha\gamma}_{\xi_1\xi_2\xi_3}
\delta(\hbar\omega+\varepsilon_{\xi_2}-\varepsilon_{\xi_1})
\right]
\end{align}}

We will eventually end up defining the quantities 
\(\Re[\beta_{xxz}]\) and \(\Im[\beta_{xyz}]\) based on the nature of the 
triple current matrix elements. In particular, the symmetry properties 
and phase structure of these matrix elements determine whether the 
corresponding nonlinear response coefficients acquire predominantly 
real or imaginary contributions. Since the three-current correlation 
involves products of interband velocity operators, different combinations 
of matrix elements can either preserve or break time-reversal-like phase 
relations, leading to distinct dispersive and absorptive parts of the 
response. As a result, \(\Re[\beta_{xxz}]\) is associated mainly with the 
reactive (dispersive) component of the nonlinear conductivity, whereas 
\(\Im[\beta_{xyz}]\) captures the dissipative or absorptive behavior. 
This separation becomes especially important when analyzing 
magneto-optical effects, where the interplay between chiral Landau levels 
and higher-energy bulk states governs the overall nonlinear optical 
response.

\subsection{Chiral-chiral contribution to the injection current}

We first analyze the contribution to the injection current arising purely from transitions between chiral Landau levels. 
These modes originate from the topologically protected zeroth Landau levels and disperse linearly along the magnetic-field direction, making them particularly relevant for nonlinear optical response\cite{Ashby2014}.

For the chiral Landau-level states described by
Eq.~(\ref{chiral_dispersion}), the current-operator matrix elements acquire a particularly simple structure defined in Eq.~(\ref{eq:current_matrix_elements}). The transverse components originate entirely
from the tilt-induced terms in the Hamiltonian, whereas the longitudinal
component is determined by the chirality-dependent group velocity along the
magnetic-field direction\cite{Zyuzin2016,Soluyanov2015}. The nonvanishing matrix elements are therefore given
by
\small{\begin{equation}
\left.
\begin{aligned}
&&  \langle n_3 | \hat{j}^x | n_1 \rangle
=\frac{\sqrt{2}\ell_B}{\hbar}t_{\parallel}
\big(
\sqrt{n_1}\,\delta_{n_3,n_1-1}
+
\sqrt{n_1+1}\,\delta_{n_3,n_1+1}
\big),
\\[0.25cm]
&&
\langle n_3 | \hat{j}^y | n_1 \rangle
=
i\frac{\sqrt{2}\ell_B}{\hbar}t_{\parallel}
\big(
\sqrt{n_1}\,\delta_{n_3,n_1-1}
-
\sqrt{n_1+1}\,\delta_{n_3,n_1+1}
\big),
\\[0.25cm]
&&
\langle n_3 | \hat{j}^z | n_1 \rangle
=
\left(t_z - v_z \eta\right)\,
\delta_{n_1,n_3}.
\end{aligned}
\right\}
\label{eq:current_matrix_elements}
\end{equation}}

Using these matrix elements, the triple-current correlators entering the
nonlinear conductivity tensor can be evaluated explicitly. One obtains
\begin{eqnarray}
\mathcal{Z}^{xxz}_{n_1 n_2 n_3}
&=&
e^3
\left(
\sqrt{2}\,t_{\parallel}\,\frac{\ell_B}{\hbar}
\right)^2
\left(t_z - v_z \eta\right)
\big[
n_1\,\delta_{n_2,n_1-1}
\nonumber\\
&&\qquad+
(n_1+1)\,\delta_{n_2,n_1+1}
\big]\delta_{n_2,n_3},
\\
\mathcal{Z}^{xyz}_{n_1 n_2 n_3}
&=&
i e^3
\left(
\sqrt{2}\,t_{\parallel}\,\frac{\ell_B}{\hbar}
\right)^2
\left(t_z - v_z \eta\right)
\big[
n_1\,\delta_{n_2,n_1-1}
\nonumber\\
&&\qquad-
(n_1+1)\,\delta_{n_2,n_1+1}
\big]\delta_{n_2,n_3}.
\end{eqnarray}

The Kronecker delta functions enforce strict dipole selection rules,
restricting optical transitions to adjacent chiral Landau levels only\cite{Ahn2017}.
Consequently, the chiral--chiral contribution to the injection current is
highly resonant and depends sensitively on both the Weyl-node chirality and the
tilt parameters. This term therefore constitutes the leading topological
component of the nonlinear magneto-optical response in the presence of a
magnetic field\cite{Juan2017,Moore2010}.

Substituting the chiral-chiral matrix elements into the general expression for the injection current tensor\cite{Sipe2000} in Eq. (\eqref{eq:beta_final}), the $\beta^{xxz}$ and $\beta^{xyz}$ component takes the form
\begin{widetext}
\begin{eqnarray}
\beta^{xxz}(\omega)
&=&
\frac{e^3\eta}{2\pi \ell_B^2 \omega^2}
\left(
t_{\parallel}\frac{\ell_B}{\hbar}
\right)^2
\sum_{n=0}^{\nu-1}
\int_{-\infty}^{\infty} dk_z
\left(t_z - v_z\eta\right)
\big\{
n
\left[
\Theta(\varepsilon_{n-1}^{\rm ch})-\Theta(\varepsilon_{n}^{\rm ch})
\right]
\left[
\delta(\varepsilon_{n-1}^{\rm ch}-\varepsilon_{n}^{\rm ch}-\omega)
+
\delta(\omega+\varepsilon_{n-1}^{\rm ch}-\varepsilon_{n}^{\rm ch})
\right]
\nonumber\\
&&\qquad\qquad
+
(n+1)
\left[
\Theta(\varepsilon_{n+1}^{\rm ch})-\Theta(\varepsilon_{n}^{\rm ch})
\right]
\left[
\delta(\varepsilon_{n+1}^{\rm ch}-\varepsilon_{n}^{\rm ch}-\omega)
+
\delta(\omega+\varepsilon_{n+1}^{\rm ch}-\varepsilon_{n}^{\rm ch})
\right]
\big\}.
\\
\beta^{xyz}(\omega)
&=&
i\frac{e^3\eta}{2\pi \ell_B^2 \omega^2}
\left(
t_{\parallel}\frac{\ell_B}{\hbar}
\right)^2
\sum_{n=0}^{\nu-1}
\int_{-\infty}^{\infty} dk_z
\left(t_z - v_z\eta\right)
\big\{
n
\left[
\Theta(\varepsilon_{n-1}^{\rm ch})-\Theta(\varepsilon_{n}^{\rm ch})
\right]
\left[
\delta(\varepsilon_{n-1}^{\rm ch}-\varepsilon_{n}^{\rm ch}-\omega)
+
\delta(\omega+\varepsilon_{n-1}^{\rm ch}-\varepsilon_{n}^{\rm ch})
\right]
\nonumber\\
&&\qquad\qquad
-
(n+1)
\left[
\Theta(\varepsilon_{n+1}^{\rm ch})-\Theta(\varepsilon_{n}^{\rm ch})
\right]
\left[
\delta(\varepsilon_{n+1}^{\rm ch}-\varepsilon_{n}^{\rm ch}-\omega)
+
\delta(\omega+\varepsilon_{n+1}^{\rm ch}-\varepsilon_{n}^{\rm ch})
\right]
\big\}.
\label{eq:beta_xxz_chiral_raw}
\end{eqnarray}

\end{widetext}

\noindent In the first term of Eq.~(\ref{eq:beta_xxz_chiral_raw}), we shift the Landau-level index according to $n \rightarrow n+1$. This allows both contributions to be combined. Also, by exploiting the linear dispersion of chiral Landau levels, the above expression simplifies to

\begin{align}
\beta^{xxz}(\omega)
&=
\frac{2\eta e^3}{4\pi \ell_B^2 \omega^2}
\left(
\sqrt{2} t_{\parallel}\frac{\ell_B}{\hbar}
\right)^2
\left(t_z - v_z\eta\right)
\sum_{n=0}^{\nu-1}(n+1)
\nonumber\\
&\quad\times
\int_{-\infty}^{\infty} dk_z\,
\Big[
\Theta(\varepsilon_n^{\rm ch})-\Theta(\varepsilon_{n+1}^{\rm ch})
\Big]
\nonumber\\
&\quad\times
\Big[
\delta(\varepsilon_n^{\rm ch}-\varepsilon_{n+1}^{\rm ch}-\omega)
+
\delta(\omega+\varepsilon_n^{\rm ch}-\varepsilon_{n+1}^{\rm ch})
\Big],
\\[0.4cm]
\beta^{xyz}(\omega)
&=
i\frac{2\eta e^3}{4\pi \ell_B^2 \omega^2}
\left(
\sqrt{2} t_{\parallel}\frac{\ell_B}{\hbar}
\right)^2
\left(t_z - v_z\eta\right)
\sum_{n=0}^{\nu-1}(n+1)
\nonumber\\
&\quad\times
\int_{-\infty}^{\infty} dk_z\,
\Big[
\Theta(\varepsilon_n^{\rm ch})-\Theta(\varepsilon_{n+1}^{\rm ch})
\Big]
\nonumber\\
&\quad\times
\Big[
\delta(\varepsilon_n^{\rm ch}-\varepsilon_{n+1}^{\rm ch}-\omega)
-
\delta(\omega+\varepsilon_n^{\rm ch}-\varepsilon_{n+1}^{\rm ch})
\Big].
\label{eq:beta_xxz_chiral_delta}
\end{align}
For chiral Landau levels, the energy difference is independent of $k_z$ and given by
$\varepsilon_{n+1}-\varepsilon_n = 2t_{\parallel}$.
Carrying out the $k_z$ integration and summation over the $m$ chiral modes, we finally obtain the closed-form expression
\begin{equation}\label{eq:beta_xxz_chiral_final}
\left.
\begin{aligned}
\Re[\beta^{xxz}(\omega)]
&=
-\frac{e^3}{\hbar}
\frac{\eta t_{\parallel}^3}{\pi \hbar^2 \omega^2}
\,\nu(\nu+1)\\&\times
\Big[
\delta(\omega-2t_{\parallel})
+
\delta(\omega+2t_{\parallel})
\Big],
\\\Im[\beta^{xyz}(\omega)]
&=
\frac{e^3}{\hbar}
\frac{\eta t_{\parallel}^3}{\pi \hbar^2 \omega^2}
\,\nu(\nu+1)\\&\times
\Big[
\delta(\omega-2t_{\parallel})
+
\delta(\omega+2t_{\parallel})
\Big].
\end{aligned}
\right\}
\end{equation}
with a condition set on $\mu$  where $\mu > 1$
Equation~(\ref{eq:beta_xxz_chiral_final}) provides a remarkably compact closed-form
expression for the purely chiral contribution to the second-order optical
conductivity tensor\cite{Bedni2024,Avdoshkin2020}. Since only transitions within the chiral Landau-level
manifold are involved, the nonlinear response is sharply localized at the
resonant frequencies \(\omega=\pm 2t_{\parallel}\), as enforced by the Dirac
delta functions. These peaks reflect the strict optical selection rules of the
chiral sector and highlight the absence of any broad continuum contribution in
this channel.
Furthermore, the prefactor scales as \(\nu(\nu+1)\), demonstrating that the
magnitude of the chiral nonlinear response is strongly enhanced for higher-order
multi-Weyl nodes. The opposite signs and the separation into
\(\Re[\beta^{xxz}]\) and \(\Im[\beta^{xyz}]\) emphasize the distinct longitudinal
and Hall-like character of the second-order response, with the imaginary part
capturing the purely antisymmetric nonlinear Hall component. Overall, this
result establishes that chiral-to-chiral transitions generate universal and
topologically amplified resonant features in the nonlinear magneto-optical
conductivity of multi-Weyl semimetals.

\subsection{Chiral-bulk contribution to the injection current}

We now consider the contribution to the nonlinear injection current arising from transitions between chiral Landau levels and bulk Landau levels\cite{Moore2010}. 
In this case, the initial or final state corresponds to a chiral Landau level, denoted by $|n\rangle$, while the intermediate state belongs to the bulk spectrum and is represented by $|n,s\rangle$, where $s=\pm$ labels the band index. 
The relevant triple current matrix element for this process is given by

\begin{align}
\mathcal{Z}^{\alpha\beta z}_{n_1 n_2 n_3}
&=
\langle n_3,s_3 | \hat{j}^{\alpha} | n_1 \rangle
\langle n_1 | \hat{j}^{\beta} | n_2,s_2 \rangle
\nonumber\\
&\quad\times
\langle n_2,s_2 | \hat{j}^{z} | n_3,s_3 \rangle .
\end{align}

We first evaluate the longitudinal current matrix element between bulk states,
\begin{align}
\langle n_2,s_2 | \hat{j}^z | n_3,s_3 \rangle
&=
\begin{pmatrix}
u_{n_2,\uparrow}^{s_2}\,\phi_{n_2-\nu} \\
u_{n_2,\downarrow}^{s_2}\,\phi_{n_2}
\end{pmatrix}^{\!\dagger}
\begin{pmatrix}
t_z+v_z\eta & 0 \\
0 & t_z-v_z\eta
\end{pmatrix}
\nonumber\\
&\quad\times
\begin{pmatrix}
u_{n_3,\uparrow}^{s_3}\,\phi_{n_3-\nu} \\
u_{n_3,\downarrow}^{s_3}\,\phi_{n_3}
\end{pmatrix}
\nonumber\\
&=
\Big[
|u_{n_2,\uparrow}^{s_2}|^2 (t_z+v_z\eta)
+
|u_{n_2,\downarrow}^{s_2}|^2 (t_z-v_z\eta)
\Big]
\nonumber\\
&\quad\times
\delta_{(n_2,s_2),(n_3,s_3)} .
\label{eq:jz_chiral_Bulk}
\end{align}
This ensures that the state \(\vert n_2,s_2\rangle= \vert n_3,s_3\rangle\) and can be written \(\vert \nu,s\rangle\) indicating a bulk state.
The relevant triple current matrix element for this process is now given by
\small{\begin{eqnarray}
\mathcal{Z}^{xx z}_{\nu-1, (\nu,s) (\nu,s)}
=
\langle \nu,s | \hat{j}^{x} | \nu-1 \rangle
\langle \nu-1 | \hat{j}^{x} | \nu,s \rangle
\langle \nu,s | \hat{j}^{z} | \nu,s \rangle \\
\mathcal{Z}^{xy z}_{\nu-1, (\nu,s) (\nu,s)}
=
\langle \nu,s | \hat{j}^{x} | \nu-1 \rangle
\langle \nu-1 | \hat{j}^{y} | \nu,s \rangle
\langle \nu,s | \hat{j}^{z} | \nu,s \rangle
\end{eqnarray}}
The \(x\) and \(y\)-components of the current matrix element between a bulk and a chiral state become
{\small
\begin{align}
\langle \nu,s | \hat{j}^x | \nu-1 \rangle
&=
\frac{\ell_B}{\hbar}
\begin{pmatrix}
u_{\nu,\uparrow}^s\,\phi_{0} \\
u_{\nu,\downarrow}^s\,\phi_\nu
\end{pmatrix}^{\!\dagger}
\begin{pmatrix}
\sqrt{2}t_{\parallel}(\hat{a}+\hat{a}^\dagger) &
\frac{\nu}{\sqrt{2}}\tilde{\lambda}_\nu(\hat{a})^{\nu-1} \\
\frac{\nu}{\sqrt{2}}\tilde{\lambda}_\nu(\hat{a}^\dagger)^{\nu-1} &
\sqrt{2}t_{\parallel}(\hat{a}+\hat{a}^\dagger)
\end{pmatrix}
\nonumber\\
&\quad\times
\begin{pmatrix}
0 \\
\phi_{\nu-1}
\end{pmatrix}
\nonumber\\[0.2cm]
&=
\frac{\ell_B}{\hbar}
\begin{pmatrix}
u_{\nu,\uparrow}^s\,\phi_{0} \\
u_{\nu,\downarrow}^s\,\phi_\nu
\end{pmatrix}^{\!\dagger}
\begin{pmatrix}
\frac{\nu}{\sqrt{2}}\tilde{\lambda}_\nu
\sqrt{(\nu-1)!}\,\phi_0 \\
\sqrt{2}t_\parallel\sqrt{\nu}\,\phi_\nu
\end{pmatrix}
\nonumber\\[0.2cm]
&=
\frac{\ell_B}{\hbar}
\Big[
u_{\nu,\uparrow}^s\,
\frac{\nu}{\sqrt{2}}\tilde{\lambda}_\nu
\sqrt{(\nu-1)!}
+
\sqrt{2}t_{\parallel}u_{\nu,\downarrow}^s\sqrt{\nu}
\Big],
\label{eq:jx_chiral_Bulk}
\end{align}}

\small{
\begin{align}
    \langle \nu,s | \hat{j}^y | \nu-1 \rangle
&=
\frac{i\ell_B}{\hbar}
\begin{pmatrix}
u_{\nu,\uparrow}^{s}\,\phi_{0} \\
u_{\nu,\downarrow}^{s}\,\phi_\nu
\end{pmatrix}^{\!\dagger}
\begin{pmatrix}
\sqrt{2}t_{\parallel}(\hat{a}+\hat{a}^\dagger) &
-\frac{\nu}{\sqrt{2}}\tilde{\lambda}_\nu(\hat{a})^{\nu-1} \\
\frac{\nu}{\sqrt{2}}\tilde{\lambda}_\nu(\hat{a}^\dagger)^{\nu-1} &
\sqrt{2}t_{\parallel}(\hat{a}+\hat{a}^\dagger)
\end{pmatrix}
\nonumber\\
&\quad\times
\begin{pmatrix}
0 \\
\phi_{\nu-1}
\end{pmatrix}
\nonumber\\
&=
\frac{i\ell_B}{\hbar}
\begin{pmatrix}
u_{\nu,\uparrow}^{s}\,\phi_{0} \\
u_{\nu,\downarrow}^{s}\,\phi_\nu
\end{pmatrix}^{\!\dagger}
\begin{pmatrix}
-\frac{\nu}{\sqrt{2}}\tilde{\lambda}_\nu
\sqrt{(\nu-1)!}\,\phi_0 \\
\sqrt{2}t_\parallel\sqrt{\nu}\,\phi_\nu
\end{pmatrix}
\nonumber\\
&=
\frac{i\ell_B}{\hbar}
\Big[
-u_{\nu,\uparrow}^{s}\,
\frac{\nu}{\sqrt{2}}\tilde{\lambda}_\nu
\sqrt{(\nu-1)!}
+
\sqrt{2}t_{\parallel}u_{\nu,\downarrow}^{s}\sqrt{\nu}
\Big].\label{eq:jy_chiral_Bulk}
\end{align}
}

Equations~(\ref{eq:jx_chiral_Bulk}) and (\ref{eq:jy_chiral_Bulk}) fully determine the chiral-bulk matrix elements entering the nonlinear conductivity. 
The resulting selection rules show that chiral-bulk transitions connect adjacent Landau levels and explicitly depend on the multi-Weyl node order through the factor $\tilde{\lambda}_\nu$.
These contributions are therefore absent in conventional Weyl semimetals ($m=1$) and constitute a distinct signature of higher-order topology.
\begin{eqnarray}\label{eq:jz_chiral_Bulk}
\langle \nu,s | \hat{j}^y | \nu-1 \rangle
=
-
\langle \nu-1 | \hat{j}^y | \nu,s \rangle^{*}.\\
\langle \nu,s | \hat{j}^x | \nu-1 \rangle
=
\langle \nu-1 | \hat{j}^x | \nu,s \rangle^{*}.
\end{eqnarray}
This antisymmetry ensures that the $xxz$ and $yyz$ components are equal,
\(
\mathcal{Z}^{xxz}_{n_1 n_2 n_3} = \mathcal{Z}^{yyz}_{n_1 n_2 n_3}.
\)
Combining Eqs.~(\ref{eq:jx_chiral_Bulk}, \ref{eq:jy_chiral_Bulk} \& \ref{eq:jz_chiral_Bulk}),
the triple current correlator 
entering the nonlinear response takes the form
\small{\begin{align}
\mathcal{Z}^{xx z}_{\nu-1, (\nu,s) (\nu,s)}
&=
\left(\frac{\ell_B}{\hbar}
\Big[
u_{\nu,\uparrow}^{s}
\frac{\nu}{\sqrt{2}}\tilde{\lambda}_\nu
\sqrt{(\nu-1)!}
+
\sqrt{2}t_{\parallel}u_{\nu,\downarrow}^{s}\sqrt{\nu}
\Big]\right)^2
\nonumber\\
&\quad\times
\Big[
|u_{\nu,\uparrow}^{s}|^2 (t_z+v_z\eta)
+
|u_{\nu,\downarrow}^{s}|^2 (t_z-v_z\eta)
\Big],
\\[0.35cm]
\mathcal{Z}^{xy z}_{\nu-1, (\nu,s) (\nu,s)}
&=
i\left(\frac{\ell_B}{\hbar}
\Big[
u_{\nu,\uparrow}^{s}
\frac{\nu}{\sqrt{2}}\tilde{\lambda}_\nu
\sqrt{(\nu-1)!}
-
\sqrt{2}t_{\parallel}u_{\nu,\downarrow}^{s}\sqrt{\nu}
\Big]\right)^2
\nonumber\\
&\quad\times
\Big[
|u_{\nu,\uparrow}^{s}|^2 (t_z+v_z\eta)
+
|u_{\nu,\downarrow}^{s}|^2 (t_z-v_z\eta)
\Big].
\end{align}}

Substituting the above Eqs. into the general expression for the injection current, Eq. (\ref{eq:beta_final}),
we obtain
\begin{widetext}
\small{\begin{equation}\label{beta_chiral_bulk}
\left.
\begin{aligned}
\beta^{xxz}(\omega)
&=
\frac{\eta}{4\pi \ell_B^2\,\omega^2}\sum_{s}
\int_{-\infty}^\infty dk_z\,
[\Theta(\varepsilon_{\nu,s}^{\rm bulk})-\Theta(\varepsilon_{\nu-1}^{\rm ch})]
\mathcal{Z}^{xx z}_{\nu-1, (\nu,s) (\nu,s)}
\left[
\delta(\varepsilon_{\nu,s}^{\rm bulk}-\varepsilon_{\nu-1}^{\rm ch}-\hbar\omega)
+
\delta(\hbar\omega+\varepsilon_{\nu,s}^{\rm bulk}-\varepsilon_{\nu-1}^{\rm ch})
\right],\\
\beta^{xyz}(\omega)
&=
\frac{i\eta}{4\pi \ell_B^2\,\omega^2}\sum_{s}
\int_{-\infty}^\infty dk_z\,
[\Theta(\varepsilon_{\nu,s}^{\rm bulk})-\Theta(\varepsilon_{\nu-1}^{\rm ch})]
\mathcal{Z}^{xy z}_{\nu-1, (\nu,s) (\nu,s)}
\left[
\delta(\varepsilon_{\nu,s}^{\rm bulk}-\varepsilon_{\nu-1}^{\rm ch}-\hbar\omega)
-
\delta(\hbar\omega+\varepsilon_{\nu,s}^{\rm bulk}-\varepsilon_{\nu-1}^{\rm ch})
\right].
\end{aligned}
\right\}
\end{equation}}
\end{widetext}

Equation~(\ref{beta_chiral_bulk}) represents the contribution to the nonlinear optical response arising from transitions between the chiral Landau level and the first bulk Landau level\cite{Tabert2016}. The difference of the step functions,
\([\Theta(\varepsilon_{\nu,s}^{\rm bulk})-\Theta(\varepsilon_{\nu-1}^{\rm ch})]\),
ensures that only optically allowed processes consistent with the occupation of initial and final states are included. The quantities \(\mathcal{Z}^{xxz}\) and \(\mathcal{Z}^{xyz}\) encode the corresponding triple current matrix elements, thereby capturing the strength of the interband coupling between chiral and bulk sectors. The Dirac delta functions enforce energy conservation for photon absorption and emission processes, with the symmetric and antisymmetric combinations distinguishing the longitudinal component \(\beta^{xxz}\) from the Hall-like component \(\beta^{xyz}\). This form highlights how the interplay between chiral and bulk Landau levels governs the frequency-dependent nonlinear conductivity in multi-Weyl semimetals.

\subsection{Bulk-bulk contribution}
With similar argument, we have, the term to consider as 
\begin{widetext}
\begin{eqnarray}
\mathcal{Z}^{xx z}_{(n_1,s_1), (n_2,s_2) (n_2,s_2)}
=
\langle n_2,s_2 | \hat{j}^{x} | n_1,s_1 \rangle
\langle n_1,s_1 | \hat{j}^{x} | n_2,s_2 \rangle
\langle n_2,s_2 | \hat{j}^{z} | n_2,s_2 \rangle \\
\mathcal{Z}^{xy z}_{(n_1,s_1), (n_2,s_2) (n_2,s_2)}
=
\langle n_2,s_2 | \hat{j}^{x} | n_1,s_1 \rangle
\langle n_1,s_1 | \hat{j}^{y} | n_2,s_2 \rangle
\langle n_2,s_2 | \hat{j}^{z} | n_2,s_2 \rangle
\end{eqnarray}
The matrix element \(\langle n_2 , s_2 | \hat{J}_x | n_1 , s_1 \rangle\) can be written as 
\begin{eqnarray}
\langle n_2 , s_2 | \hat{J}_x | n_1 , s_1 \rangle
&=&
e \left(\frac{\ell_B}{\hbar}\right)
\left\{
\mathcal{M}_{n_1,n_2,s_1,s_2}\,
\delta_{n_2,n_1-1}+\mathcal{M}_{n_1-1,n_2,s_1,s_2}\,
\delta_{n_2,n_1-1}\right\}
\end{eqnarray}
where we have defined \(\mathcal{M}_{n_1,n_2,s_1,s_2}=\big[
\sqrt{2}\, t_\parallel \,
u_{n_1,\uparrow}^{s_1}u_{n_2,\uparrow}^{s_2}\sqrt{n_1-\nu+1}
+
\frac{\nu}{\sqrt{2}}\,
\tilde{\lambda}_\nu\,
u_{n_1,\downarrow}^{s_1}u_{n_2,\uparrow}^{s_2}
\sqrt{\frac{n_1!}{(n_1-\nu+1)!}}
+
\sqrt{2}\, t_\parallel \,
u_{n_1,\downarrow}^{s_1}u_{n_2,\downarrow}^{s_2}\sqrt{n_1+1}
\big]\). 
Analogously, we can writre 
\begin{eqnarray}
\langle n_1 , s_1 | \hat{J}^{y} | n_3 , s_2 \rangle
&=&
i e \left(\frac{\ell_B}{\hbar}\right)
\left\{
-\mathcal{M}_{n_1,n_2,s_1,s_2}\,
\delta_{n_2,n_1-1}+\mathcal{M}_{n_1-1,n_2,s_1,s_2}\,
\delta_{n_2,n_1-1}\right\}
\end{eqnarray}

\noindent After enforcing the selection rules, the $xxz$ and $yyz$ tensor components are equal and can be written as
\begin{eqnarray}\label{Z_xxz}
&&\mathcal{Z}^{xx z}_{(n_1,s_1), (n_2,s_2) (n_2,s_2)}
= e^3 \left(\frac{\ell_B}{\hbar}\right)^2
\left\{
\mathcal{M}_{n_1,n_2,s_1,s_2}\,
\delta_{n_2,n_1-1}+\mathcal{M}_{n_1-1,n_2,s_1,s_2}\,
\delta_{n_2,n_1+1}\right\}\nonumber\\&&\qquad\times
\left\{
\mathcal{M}_{n_2,n_1,s_2,s_1}\,
\delta_{n_1,n_2-1}+\mathcal{M}_{n_2-1,n_1,s_2,s_1}\,
\delta_{n_1,n_2+1}\right\}
\Big[
|u_{n_2,\uparrow}^{s_2}|^2 (t_z+v_z\eta)
+
|u_{n_2,\downarrow}^{s_2}|^2 (t_z-v_z\eta)
\Big]
\delta_{(n_2,s_2),(n_3,s_3)} .
\nonumber\\&&\qquad
= e^3 \left(\frac{\ell_B}{\hbar}\right)^2
\left\{
(\mathcal{M}_{n_1,n_2,s_1,s_2})^2\,
\delta_{n_2,n_1-1}+(\mathcal{M}_{n_1-1,n_2,s_1,s_2})^2\,
\delta_{n_2,n_1+1}\right\}
\left[
|u_{n_2,\uparrow}^{s_2}|^2 (t_z+v_z\eta)
+
|u_{n_2,\downarrow}^{s_2}|^2 (t_z-v_z\eta)
\right].
\end{eqnarray}
\begin{eqnarray}\label{Z_xyz}
&&\mathcal{Z}^{xy z}_{(n_1,s_1), (n_2,s_2) (n_2,s_2)}
=i e^3 \left(\frac{\ell_B}{\hbar}\right)^2
\left\{
(\mathcal{M}_{n_1,n_2,s_1,s_2})\,
\delta_{n_2,n_1-1}+\mathcal{M}_{n_1-1,n_2,s_1,s_2}\,
\delta_{n_2,n_1+1}\right\}\nonumber\\&&\qquad\times
\left\{-
\mathcal{M}_{n_2,n_1,s_2,s_1}\,
\delta_{n_1,n_2-1}+\mathcal{M}_{n_2-1,n_1,s_2,s_1}\,
\delta_{n_1,n_2+1}\right\}
\left[
|u_{n_2,\uparrow}^{s_2}|^2 (t_z+v_z\eta)
+
|u_{n_2,\downarrow}^{s_2}|^2 (t_z-v_z\eta)
\right]
\delta_{(n_2,s_2),(n_3,s_3)} \nonumber\\
&&\qquad=i e^3 \left(\frac{\ell_B}{\hbar}\right)^2
\left\{
(\mathcal{M}_{n_1-1,n_2,s_1,s_2})^2\,
\delta_{n_2,n_1+1}-(\mathcal{M}_{n_1,n_2,s_1,s_2})^2\,
\delta_{n_2,n_1-1}\right\}
\left[
|u_{n_2,\uparrow}^{s_2}|^2 (t_z+v_z\eta)
+
|u_{n_2,\downarrow}^{s_2}|^2 (t_z-v_z\eta)
\right]
\end{eqnarray}

Substituting Eqs.~(\ref{Z_xxz}) and (\ref{Z_xyz}) into the general expression for the second-order response yields
\begin{equation}\label{beta_bulk}
\left.
\begin{aligned}
\beta^{xxz}
&= \frac{\eta e^{3}}{4\pi \ell_B^{2}\omega^{2}} v_z
\left( \frac{\ell_B}{\hbar} \right)^{2}
\sum_{\substack{n\\ s\neq s'}}
\int_{-\infty}^{\infty} dk_z \,
\Big[
\Theta\!\left(\varepsilon_{n+1,s^\prime}^{\rm bulk}\right)
- \Theta\!\left(\varepsilon_{n,s}^{\rm bulk}\right)
\Big]\mathcal{M}_{n,n+1,s,s^\prime}
\\
&\quad \times
\Big(
|u_{n+1,\uparrow}^{s'}|^{2}
- |u_{n,\uparrow}^{s}|^{2}
- |u_{n+1,\downarrow}^{s'}|^{2}
+ |u_{n,\downarrow}^{s}|^{2}
\Big)
\Big\{
\delta\!\left(-\varepsilon_{n,s}^{\rm bulk}+\varepsilon_{n+1,s^\prime}^{\rm bulk}+\omega\right)
+
\delta\!\left(-\omega-(\varepsilon_{n,s}^{\rm bulk}-\varepsilon_{n+1,s^\prime}^{\rm bulk})\right)
\Big\},
\\
\beta^{xyz}
&= i\frac{\eta e^{3}}{4\pi \ell_B^{2}\omega^{2}} v_z
\left( \frac{\ell_B}{\hbar} \right)^{2}
\sum_{\substack{n\\ s\neq s'}}
\int_{-\infty}^{\infty} dk_z \,
\Big[
\Theta\!\left(\varepsilon_{n+1,s^\prime}^{\rm bulk}\right)
- \Theta\!\left(\varepsilon_{n,s}^{\rm bulk}\right)
\Big]\mathcal{M}_{n,n+1,s,s^\prime}
\\
&\quad \times
\Big(
|u_{n+1,\uparrow}^{s'}|^{2}
- |u_{n,\uparrow}^{s}|^{2}
- |u_{n+1,\downarrow}^{s'}|^{2}
+ |u_{n,\downarrow}^{s}|^{2}
\Big)
\Big\{
\delta\!\left(-\varepsilon_{n,s}^{\rm bulk}+\varepsilon_{n+1,s^\prime}^{\rm bulk}+\omega\right)
-
\delta\!\left(-\omega-(\varepsilon_{n,s}^{\rm bulk}-\varepsilon_{n+1,s^\prime}^{\rm bulk})\right)
\Big\}.
\end{aligned}
\right\}
\end{equation}

   \begin{figure*}
     \centering
       \includegraphics[width=190mm,height=70.5mm]{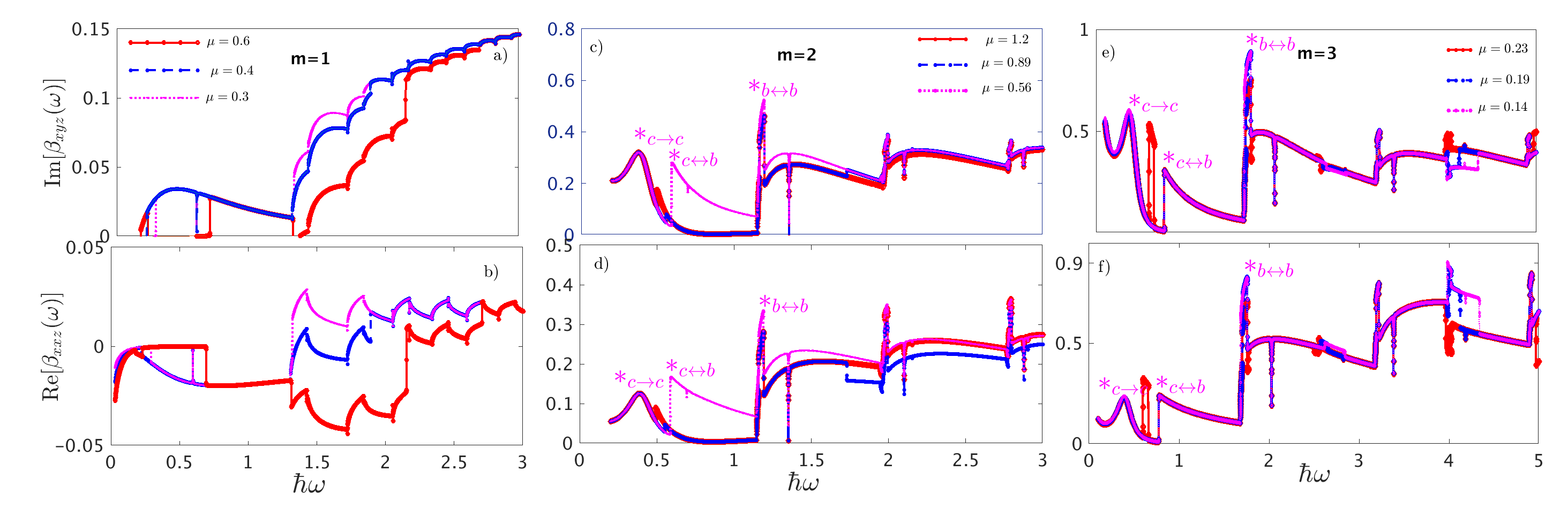}

    \caption{The components of second-order DC conductivities $\beta^{xyz}$ and  $\beta^{xxz}$( in the units of $\frac{e^3}{\hbar}$) as a function of the photon energy{ $\hbar\omega$}. The plots are obtained  for multi-Weyl nodes with finite tilting parameters($t_{\parallel},t_z \ne 0$) where chiral to chiral transitions becomes significant.The various curves illustrate the behaviour at different chemical potential.}
     \label{fig:beta_m123}
 \end{figure*}
 \end{widetext}
Using the analytical expressions derived in the previous sections, we now
present the full frequency-dependent nonlinear response of multi-Weyl
semimetals in the presence of a perpendicular magnetic field.
Figure~\ref{fig:beta_m123} displays the behavior of the Hall-like component
\(\Im[\beta^{xyz}(\omega)]\) and the longitudinal component
\(\Re[\beta^{xxz}(\omega)]\) for different winding numbers
\(\nu=1,2,\) and \(3\), and for several values of the chemical potential \(\mu\).

The total nonlinear conductivity is obtained by incorporating all relevant
Landau-level transition channels\cite{König2017,Tabert2016,Ahn2017,Weng2015,
Ma2019, Lu2015, Burkov2015,Goerbig2017,  Behrends2019, João2019, Fu2022}. In particular, the contribution arising
purely from chiral Landau-level transitions is governed by the compact
closed-form result in Eq.~(\ref{eq:beta_xxz_chiral_final}), which produces
distinct low-energy resonances associated with the chiral sector.
In addition, mixed transitions between the chiral level and higher bulk
Landau levels are captured by Eq.~(\ref{beta_chiral_bulk}), leading to
additional peaks and threshold structures in the intermediate-frequency
regime. Finally, the dominant high-energy response originates from bulk-to-bulk
interband processes described by Eq.~(\ref{beta_bulk}),
which become increasingly prominent as more bulk Landau levels participate in
optical transitions.

As evident from Fig.~\ref{fig:beta_m123}, the nonlinear response exhibits a
series of sharp resonant features and step-like variations, reflecting the
energy-conserving delta-function selection rules imposed by Landau-level
quantization. The onset of the spectra at low photon energies is primarily
dictated by chiral-to-chiral (\(c\!\to\!c\)) transitions, which provide the
leading contribution in the vicinity of the chiral Landau level.
With increasing frequency, additional absorption thresholds emerge from
chiral-to-bulk (\(c\!\to\!b\)) processes, marking the activation of transitions
between the topologically protected chiral mode and the first bulk Landau
levels. At still higher energies, the response becomes dominated by bulk-to-bulk
(\(b\!\to\!b\)) interband transitions, resulting in multiple resonant peaks as
successive bulk Landau levels enter the optical window.

Moreover, the overall magnitude and complexity of the spectra increase
systematically with the winding number \(\nu\), demonstrating that higher-order
Weyl nodes significantly enhance the nonlinear magneto-optical conductivity
through the interplay of chiral and bulk excitations.

\subsection{Nonlinear Magneto-Optical Response in the Untilting Limit}

In this section we evaluate the second-order optical conductivity tensor for
multi-Weyl semimetals in the absence of tilting. We set the in-plane and
out-of-plane tilt parameters to zero, i.e., $t_{\parallel}=0$ and $w_{z}=0$,
so that the Landau-level spectrum remains fully symmetric about the Weyl node.
This limit is particularly important because it allows the nonlinear optical
tensor to be expressed in a closed and compact analytical form. In the presence
of tilt, the Landau-level structure becomes asymmetric and the resulting
nonlinear response generally involves complicated energy dispersions, making it
difficult to obtain transparent expressions beyond numerical evaluation.

By focusing on the untilted case, we are able to isolate the intrinsic
topological contribution of multi-Weyl nodes and derive simplified results for
the second-order conductivity components in terms of a few universal parameters.
Such compact formulas provide direct physical insight into the role of chiral
Landau levels and optical selection rules, and serve as a useful reference
point for understanding how additional effects, such as tilting or disorder,
modify the nonlinear magneto-optical response.
\begin{figure*}[t]
        \centering
            \includegraphics[width=190mm,height=70.5mm]{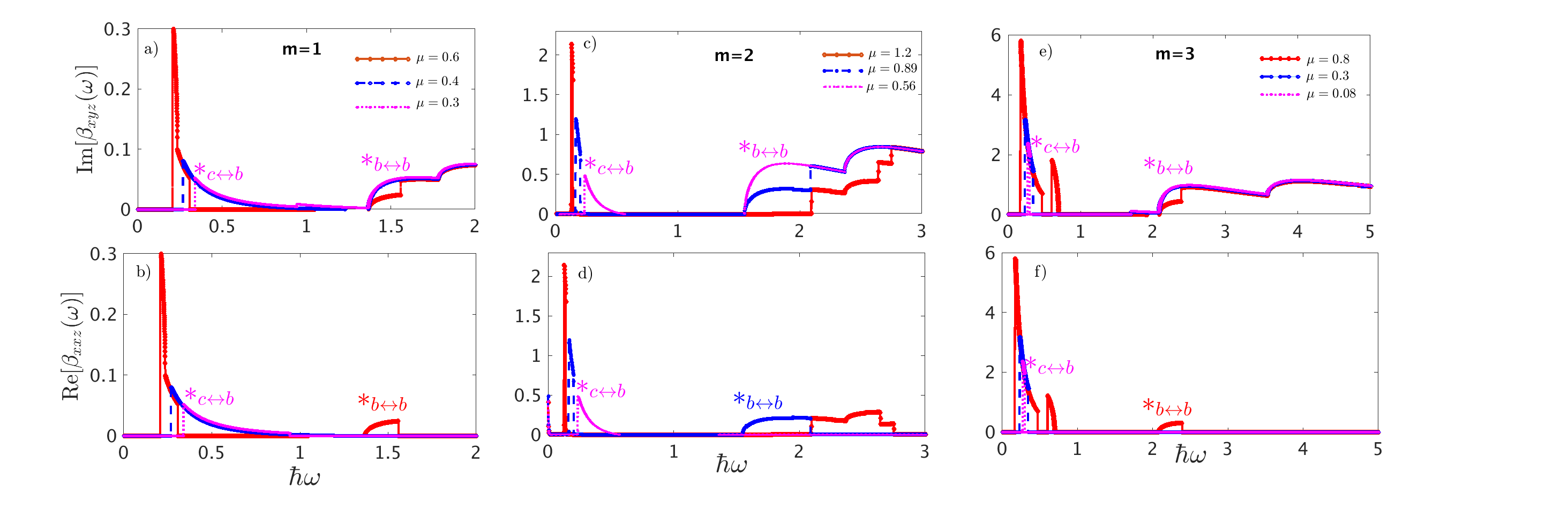}
             \caption{ The photon-energy dependence of the second-order DC conductivity components
$\beta^{xyz}$ and $\beta^{xxz}$ (in units of $e^3/\hbar$) is presented for
multi-Weyl nodes in the untilted limit ($t_{\parallel}=0$, $t_z=0$), as
derived from Eqs.~(\ref{beta_untilted_bulk}) and~(\ref{chiral_bulk_analytic}).
The individual curves correspond to different chemical potentials.
}
        \label{fig:placeholder}
    \end{figure*}
    
\noindent{\textbf{Bulk-bulk nonlinear response in the untilting limit}}: 
In this limit, the Landau-level spectrum becomes symmetric about the Weyl node,
and the bulk Landau-level energies reduce to
$s$-dependent branches of the form
$\varepsilon_{n,s}^{\rm bulk}(k_z)=s\,\Gamma_n^\nu(k_z)$.
As a result, the band index $s=\pm$ directly labels electron- and hole-like
states, and optical transitions contributing to the nonlinear response are
restricted to interband processes with $s\neq s'$.
The absence of tilt considerably simplifies the structure of the current matrix
elements.
In particular, the difference of spinor weights appearing in
Eq.~(\ref{beta_bulk}),
\(
|u_{n+1,\uparrow}^{s'}|^{2}
- |u_{n,\uparrow}^{s}|^{2}
- |u_{n+1,\downarrow}^{s'}|^{2}
+ |u_{n,\downarrow}^{s}|^{2},
\)
can be expressed solely in terms of the Landau-level energies
$\Gamma_n^\nu(k_z)$ and $\Gamma_{n+1}^\nu(k_z)$.
This cancellation reflects the restoration of particle-hole symmetry in the
untilted Hamiltonian and ensures that only interband coherence contributes to
the second-order optical response\cite{Juan2017,Fang2012}.
The delta functions in Eq.~(\ref{beta_bulk}) enforce energy
conservation for optical transitions between adjacent bulk Landau levels,
\(
\varepsilon_{n+1,s'}^{\rm bulk}(k_z)
-
\varepsilon_{n,s}^{\rm bulk}(k_z)
=
\pm \hbar\omega .
\)
Since $\Gamma_n^\nu(k_z)$ depends quadratically on $k_z$, these constraints define
two symmetric roots in $k_z$, which allows the momentum integral to be evaluated
analytically.
Carrying out the $k_z$ integration converts the delta functions into inverse
Jacobian factors, producing the characteristic square-root structure associated
with the one-dimensional joint density of states along the magnetic-field
direction.
The difference between the longitudinal component $\beta^{xxz}$ and the
Hall-type component $\beta^{xyz}$ arises entirely from the relative sign between
the two delta-function contributions in Eq.~(\ref{beta_bulk}).
While both components share the same resonance structure, this sign difference
leads to qualitatively distinct frequency dependences:
$\beta^{xyz}$ acquires an imaginary prefactor and is odd under time-reversal,
whereas $\beta^{xxz}$ remains purely real.
Collecting all contributions, we arrive at the closed-form expressions for the
nonlinear bulk-bulk response in the untilting limit, presented below.

\begin{widetext}

\begin{equation}\label{beta_untilted_bulk}
\left.
\begin{aligned}
\beta^{xyz}(\omega)
&=
i\,\frac{e^3}{\hbar}
\frac{\varepsilon_B^2 f^2(\nu,n)}
{8\pi\!\left[\omega^4-\big(\varepsilon_B f(\nu,n)\big)^4\right]}
\sum_{n}
\Big[
\Theta\!\left(\omega^2+f^2(\nu,n)\varepsilon_B^2-|2\mu\omega|\right)
\\
&\qquad
+
\operatorname{sgn}\!\Big[
\Theta\!\left(\omega^2-f^2(\nu,n)\varepsilon_B^2-|2\mu\omega|\right)
\Big]
\sqrt{
f^4(\nu,n)\varepsilon_B^4+(\hbar\omega)^4
-2(\hbar\omega)^2\varepsilon_B^2 g^2(\nu,n)
}
\Big],
\\[0.4cm]
\beta^{xxz}(\omega)
&=
\frac{e^3}{\hbar}
\frac{\varepsilon_B^2 f^2(\nu,n)}
{8\pi\!\left[\omega^4-\big(\varepsilon_B f(\nu,n)\big)^4\right]}
\sum_{n}
\Big[
\Theta\!\left(\omega^2+f^2(\nu,n)\varepsilon_B^2-|2\mu\omega|\right)
\\
&\qquad
-
\Theta\!\left(\omega^2-f^2(\nu,n)\varepsilon_B^2-|2\mu\omega|\right)
\sqrt{
f^4(\nu,n)\varepsilon_B^4+(\hbar\omega)^4
-2(\hbar\omega)^2\varepsilon_B^2 g^2(\nu,n)
}
\Big].
\end{aligned}
\right\}
\end{equation}

\end{widetext}

\noindent where the structure functions $f(\nu,n)$ and $g(\nu,n)$ are defined as
\begin{eqnarray}
g^2(\nu,n)=\frac{2(n+1)-\nu}{n-\nu+1}\,
\frac{n!}{(n-\nu)!},
\\
f^2(\nu,n)=\frac{-\nu\,n!}{(n+1-\nu)(n-\nu)!}    
\end{eqnarray}

\noindent These functions encode the dependence of the nonlinear response on the
topological charge of the Weyl node and the Landau-level index.  The step
functions $\Theta$ restrict the transitions to those satisfying the optical
selection rules, while the square-root term represents the joint density of
states for allowed transitions.

\vspace{0.5em}
\noindent
\textit{\textbf{Chiral to bulk transitions.}}
In the absence of tilt, the longitudinal current matrix element entering
$\mathcal{Z}^{\alpha\beta z}_{\nu-1,(\nu,s)(\nu,s)}$ reduces to a purely
velocity-dependent factor, while the transverse matrix elements depend only on
the multi-Weyl coupling $\tilde{\lambda}_\nu$.
Consequently, the triple current correlator factorizes into an energy-dependent
prefactor multiplied by a monopole-charge-dependent constant.
This allows the chiral-bulk response to be expressed entirely in terms of the
magnetic energy scale $\varepsilon_B$ and the monopole charge $\nu$.

The delta functions in Eq.~(\ref{beta_chiral_bulk}) enforce energy
conservation for optical transitions between the chiral and bulk Landau levels,
\[
\varepsilon_{\nu,s}^{\rm bulk}(k_z)
-
\varepsilon_{\nu-1}^{\rm ch}(k_z)
=
\pm \hbar\omega .
\]
Since $\Gamma_\nu^\nu(k_z)$ depends quadratically on $k_z$, this constraint
defines symmetric roots in momentum space, allowing the $k_z$ integration to be
performed analytically.
Carrying out the integration converts the delta functions into inverse Jacobian
factors, producing a characteristic rational frequency dependence that is
distinct from the bulk-bulk response.

After evaluating the $k_z$ integral and summing over the bulk band index $s$,
the nonlinear response assumes a universal form governed by a single structure
function $f(\nu)=\nu!$.
The difference between the longitudinal component $\beta^{xxz}$ and the
Hall-type component $\beta^{xyz}$ arises from the relative sign between the two
delta-function contributions in Eq.~(\ref{beta_chiral_bulk}).
While both components share identical resonance conditions, $\beta^{xyz}$
acquires an imaginary prefactor and is odd under time-reversal symmetry, whereas
$\beta^{xxz}$ remains purely real.

Collecting all contributions, we obtain the closed-form expressions for the
chiral-bulk nonlinear injection current in the untilting limit. 
These results reveal a pronounced low-frequency enhancement controlled by the
topological chiral Landau level and provide a direct nonlinear optical signature
of the multi-Weyl monopole charge.
\begin{widetext}
\begin{eqnarray}\label{chiral_bulk_analytic}
\left.
\begin{aligned}
\beta^{xyz}(\omega)
=
i\,\frac{e^3}{\hbar}
\frac{\varepsilon_B^2 f^2(\nu)\,\nu}
{16\pi\omega^2}
\Bigg[
\Theta\!\left(\omega^2+f^2(\nu)\varepsilon_B^2-|2\mu\omega|\right)
+
\operatorname{sgn}\!\Big[
\Theta\!\left(\omega^2-f^2(\nu)\varepsilon_B^2-|2\mu\omega|\right)
\Big]\,
\frac{-\omega^2+f^2(\nu)\varepsilon_B^2\nu}
{\omega^2+f^2(\nu)\varepsilon_B^2\nu}
\Bigg],
\\
\beta^{xxz}(\omega)
=
\frac{e^3}{\hbar}
\frac{\varepsilon_B^2 f^2(\nu)\,\nu}
{16\pi\omega^2}
\Bigg[
\Theta\!\left(\omega^2+f^2(\nu)\varepsilon_B^2-|2\mu\omega|\right)-
\Theta\!\left(\omega^2-f^2(\nu)\varepsilon_B^2-|2\mu\omega|\right)
\frac{-\omega^2+f^2(\nu)\varepsilon_B^2\nu}
{\omega^2+f^2(\nu)\varepsilon_B^2\nu}
\Bigg],
\end{aligned}
\right\}
\end{eqnarray}
\end{widetext}
with $f^2(\nu)=\nu!$.
The chiral-bulk contribution is particularly significant at low frequencies,
since it directly reflects the topological character of the chiral Landau
level and its coupling to the electromagnetic field.

\begin{figure}[t]
    \centering
    \includegraphics[width=\linewidth]{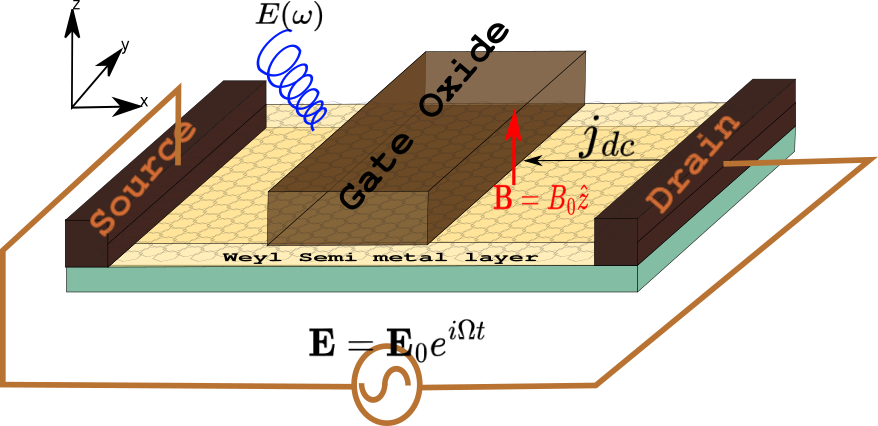}
    \caption{
    Schematic of a gated transport geometry for detecting nonlinear
    magneto-optical injection currents in a multi-Weyl semimetal.
    A multi-Weyl channel is contacted by metallic source and drain
    electrodes and electrostatically tuned through a gate oxide layer.
    A perpendicular magnetic field $\mathbf{B}=B_{0}\hat{z}$ quantizes
    the spectrum into chiral and bulk Landau levels, while illumination
    by a circularly polarized optical field $E(\omega)=E_{0}e^{i\Omega t}$
    drives resonant inter-Landau level transitions.
    The resulting asymmetric carrier population generates a dc
    photocurrent $j_{\mathrm{dc}}$ along the channel, providing a direct
    transport probe of the second-order injection tensor
    $\beta^{\alpha\beta\gamma}(\omega)$ and its characteristic
    chiral-chiral, chiral-bulk, and bulk-bulk contributions discussed in
    the paper.
    }
    \label{fig:device}
\end{figure}

\paragraph*{\textbf{Experimental relevance and gated device geometry:}}
The nonlinear magneto-optical injection current derived in this work can
be probed in a gated transport configuration closely analogous to a
field-effect transistor geometry, where a multi-Weyl semimetal forms the
conducting channel between metallic source and drain contacts and is
electrostatically tuned by a gate electrode separated through an
insulating oxide layer. In the presence of a perpendicular magnetic
field $\mathbf{B}=B_{0}\hat{z}$, the electronic spectrum is quantized into
bulk and chiral Landau levels, providing the microscopic origin of the
frequency-selective resonances identified in the second-order
conductivity tensor. Illumination by a circularly polarized optical
field $E(\omega)=E_{0}e^{i\Omega t}$ drives resonant inter-Landau-level
transitions whose asymmetric population dynamics generate a dc
photocurrent $j_{\mathrm{dc}}$ along the channel via the injection
mechanism captured by the tensor $\beta^{\alpha\beta\gamma}(\omega)$\cite{Zeng2021}.
Within this geometry, the gate voltage directly controls the chemical
potential relative to the chiral and bulk Landau levels, enabling
experimental isolation of the distinct chiral-chiral, chiral-bulk, and
bulk-bulk contributions predicted by the theory, as well as their
characteristic monopole-charge scaling and tilt-dependent spectral
structure. The resulting source-drain photocurrent therefore provides a
direct transport signature of the nonlinear magneto-optical response and
offers a realistic route for detecting topological enhancement of
injection currents in multi-Weyl semimetals.

\section{Conclusions}\label{conclusions}

In this work, we developed a comprehensive microscopic theory of nonlinear
magneto-optical injection currents in multi-Weyl semimetals subjected to a
uniform magnetic field. Starting from a tilted multi-Weyl Hamiltonian with
arbitrary monopole charge $\nu$, we derived the full Landau-level spectrum and
formulated the second-order conductivity tensor within a Kubo-type nonlinear
response framework. This approach enabled a unified treatment of optical
transitions involving chiral, mixed chiral-bulk, and bulk Landau levels,
thereby revealing the complete structure of the nonlinear magneto-optical
response.

Our analysis shows that transitions confined within the chiral Landau-level
manifold generate universal resonant peaks sharply localized at frequencies
determined solely by the in-plane tilt parameter. The magnitude of this purely
topological contribution scales as $\nu(\nu+1)$, demonstrating a strong
enhancement of nonlinear response with increasing Weyl-node order. Mixed
chiral-bulk transitions introduce additional intermediate-frequency structures
governed by the magnetic energy scale and monopole charge, while bulk-to-bulk
processes dominate the high-frequency regime through a sequence of Landau-level
resonances and threshold features. Together, these channels produce a
characteristic multiscale spectral profile that directly encodes the topology of
multi-Weyl nodes.

In the absence of tilt, we obtained compact closed-form analytical expressions
for the nonlinear conductivity tensor. These results isolate the intrinsic
topological contribution of multi-Weyl semimetals and reveal universal scaling
governed by the magnetic energy scale and factorial structure functions
associated with higher monopole charge. The distinct behaviors of the
longitudinal component $\Re[\beta^{xxz}]$ and the Hall-type component
$\Im[\beta^{xyz}]$ further clarify the separation between dispersive and
absorptive nonlinear magneto-optical responses.

Overall, our findings establish nonlinear magneto-optical injection currents as
a sensitive probe of higher-order Weyl topology and chiral Landau physics. The
predicted resonant structures, monopole-charge scaling, and universal frequency
dependences provide clear experimental signatures accessible through terahertz
or infrared spectroscopy in candidate multi-Weyl materials. More broadly, this
work highlights the power of nonlinear optical responses in strong magnetic
fields as a route toward identifying and characterizing topological quantum
matter beyond the linear-response paradigm.

Future directions include the incorporation of disorder and finite relaxation
dynamics, extension to finite temperature and interaction effects, and
exploration of nonlinear responses in related topological systems such as
multifold fermions and pseudospin-1 semimetals.


\end{document}